\UseRawInputEncoding

\documentclass[conference]{IEEEtran}

\IEEEoverridecommandlockouts
\usepackage{cite}
\usepackage{amsmath,amsfonts,amsthm,bm,amssymb} %

\usepackage{algorithm}
\usepackage{algpseudocode} 
\usepackage{graphicx}
\usepackage{graphics}
\usepackage{epsfig}
\usepackage{footnote}
\usepackage{textcomp}
\PassOptionsToPackage{no-math}{fontspec}
\usepackage{xcolor}
\usepackage{tikz}  
\usepackage{makecell}
\usepackage{bm}
\usetikzlibrary{calc} 
\usetikzlibrary{arrows,shapes,chains} 
\usetikzlibrary{positioning, shapes.geometric}
\usepackage{multirow} 
\usepackage{verbatim}
\usepackage{enumitem}
\def\BibTeX{{\rm B\kern-.05em{\sc i\kern-.025em b}\kern-.08em
		T\kern-.1667em\lower.7ex\hbox{E}\kern-.125emX}}
	
\renewcommand{\algorithmicrequire}{ \textbf{Input:}}      
\renewcommand{\algorithmicensure}{ \textbf{Output:}}     
\usepackage{verbatim}
\usepackage{pgfplots}
\usepackage{hyperref}
\pgfplotsset{width=10cm,compat=1.9}
\pgfplotsset{
	every axis legend/.style={
		cells={anchor=center},
		inner xsep=3pt,inner ysep=2pt,
		nodes={inner sep=2pt,text depth=0.1em},
		anchor=north east,
		shape=rectangle,
		fill=white,draw=green,
		font=\footnotesize
	},
}
\def\pgfsysdriver{pgfsys -dvipdfmx.def}
\usepackage{verbatim}
\usepackage{bm}
\normalsize
	
\graphicspath{{./figs/}}
\begin{document}
\bibliographystyle{unsrt}
\title{A recipe of training neural network-based LDPC decoders\\
}

\author{Guangwen Li, Xiao Yu

\thanks{G.Li is with the College of Information \& Electronics, Shandong Technology and Business University, Yantai, China e-mail: lgwa@sdu.edu.cn}
\thanks{X.Yu is with the Department of Physical Sports, Binzhou Medical University, Yantai, China e-mail: yuxiao@bzmu.edu.cn}
}

\maketitle

\begin{abstract}
	It is known belief propagation decoding variants of LDPC codes can be unrolled easily as neural networks after assigning differed weights to message passing edges flexibly. In this paper we focus on how to determine these weights, in the form of trainable parameters, within a framework of deep learning. Firstly, a new method is proposed to generate high-quality training data via exploiting an approximation to the targeted mixture density. Then the strong positive correlation between training loss and decoding metrics is fully exposed after tracing the training evolution curves. Lastly, for the purpose of facilitating training convergence and reducing decoding complexity, we highlight the necessity of slashing the number of trainable parameters while emphasizing the locations of these survived ones, which is justified in the extensive simulation. 
\end{abstract}


\section{Introduction}
In modern telecommunication system, the error correction codes, as an indispensable ingredient, appears to combat unavoidable random noises in the communication channel via deliberately designed redundancy in encoding. Among them, the family of linear low-density parity-check codes (LDPC) \cite{Gallager1962} demonstrate  prominent error correction capability by achieving near Shannon Limit asymptotically. However, the accompanied iterative belief propagation (BP) decoders  require substantial computational resources, thus unbearable in some harsh applications.  With the target of reducing decoding complexity effectively, the min-sum (MS) and its variants, normalized min-sum  and offset min-sum (OMS) (NMS) \cite{Zhao2005,Jiang2006} were proposed as an approximation to the BP. Regretfully, the incurred performance loss is not negligible in most cases. 

On the other hand, with the giant leap of computational power and rapid development of machine learning theory, various deep learning strategies were deployed successfully in the fields from traditional image processing  \cite{Razzak2018}, objection detection \cite{Wang2018}, face recognition  \cite{hu2015face} to more challenging tasks such as natural language processing  \cite{Young2018} and autonomous driving \cite{Grigorescu2020} etc. Their success is symbolized finally with the state-of-the-art performance.

Many attempts have recently been explored to integrate deep learning methods into the realm of LDPC codes decoding. To overcome the challenging obstacle in code space, known as curse of dimensionality \cite{gruber2017deep}, the authors in  \cite{Lugosch2017, Nachmani2018, Nachmani2016, Liang2018, Lugosch2018b, Wang2020} proposed the neural normalized min-sum (NNMS) or neural offset min-sum (NOMS) decoders, after unrolling the original min-sum decoding schemes  into typical neutral networks such as convolutional neural network (CNN) or recurrent neural network (RNN) with respect to the Tanner graph of a code in lieu of an arbitrary neural network, It was verified that such approaches can achieve near maximum a posteriori (MAP) decoding performance for classical codes such as BCH codes and short Gallager LDPC codes \cite{Helmling2019}. Furthermore, for a class of protograph-based 5G LDPC codes of long block, whose structure is quasi cyclic,  \cite{Dai2021, Wu2018} explored to apply machine learning techniques with fully sharing edge weights and achieved better decoding performance. Besides that, for a class of irregular LDPC codes, it was proposed in  \cite{Wang2021} to assign and share weights  for the NNMS decoder in terms of the degree distribution of the code, thus outperformed its counterparts.

Sticking to the design of NNMS decoder, we elaborate on how to optimize its training to obtain a universal decoder with excellent performance in a wide range of signal noise ratio (SNR) region. Our main contribution consists of one innovation and two findings. Since the feeding of a neural decoder impacts substantially its training outcome, we propose to generate input data  with respect to a novel approximation to the targeted mixture probability density. Then the relation between training loss and decoding metrics is clarified statistically, by way of explaining the underlying changes along with the training iteration of all evolution curves. Lastly, it is found the performance of a neural decoder is greatly affected by the locations of its trainable parameters, instead of their amount. Therefore, with a proper configuration of trainable parameters, the benefits we gain range from lightening the training burden to reducing the complexity of the finalized decoder, which is verified in the followed simulations.

The remainder of the paper  is organized as follows. Section \ref{background} presents the necessary background about LDPC decoding variants, as well as the neural networking structure of NNMS. The motivation of our scheme is presented in Section \ref{motivation}, while experimental results  are discussed in detail in Section \ref{simulations}.  Section \ref{conclusions} concludes this work with some remarks and suggestions of further research directions.

\section{Background}
\label{background}
For error correction codes, the redundancy added in in channel encoding makes it possible to transmit information reliably over unreliable channels. 

Assume each code bit ${c}_{i},i=1,2, \cdots \in \mathit{N}$ is modulated with BPSK via $1-2{c}_{i}$, where $\mathit{N}$ denotes code block length. Due to the disturbance of additive white Gaussian noise (AWGN) of zero mean and variance $\sigma^2$, the log-likelihood ratio (LLR) of the $i$-th bit  at channel output is given by
\begin{equation}
{b_{{v_i}}} = \log \left( {\frac{{p({y_i}|{{{c}}_i} = 0}}{{p({y_i}|{{{c}}_i} = 1}}} \right) = \frac{{2{y_i}}}{{{\sigma ^2}}}
\end{equation} where $y_{i}$ is the $i$-th noisy signal. As a result, a positive(negative) $y_{i}$  hints the bit sent being ‘0’('1'). Then these LLRs will be regarded as channel messages involved in the BP.

\subsection{Standard BP and some MS variants}
 The bipartite Tanner graph of a code, whose structure pertains totally to its parity check matrix $\bm{H}$, consists of  $\mathit{N}$ variable nodes and  $\mathit{M}$ check nodes. And each '1' at the $i$-th row  and the $j$-th column of $\mathit{N}$  denotes variable node $i$ and check node $j$ is connected with an edge in the Tanner graph. 

For LDPC codes, the standard BP is a competitive one among many decoding schemes, in the sense it is optimal under the assumption of a tree-like Tanner graph without any cycles. However, it degrades to being suboptimal for any code of practical implementation, since the unavoidable cycles of the Tanner graph  make the passed messages between nodes in the BP no longer independent of each other.

To be self-contained, let us take a close look at the decoding process of BP. For $i\in\{1,2,...,N\}$, $j\in\{1,2,...,M\}$, $l\in\{1,2,...,T\}$, where $T$ is the maximum number of iterations, at the $\mathit{l}$-th iteration, the message from variable node $v_i$ to  check node $c_j$ is \ref{eq_v2c}

\begin{equation}
x_{v_i \to c_j}^{(l)}  = {b_{v_i}} + \sum\limits_{\substack{c_p \to v_i\\p \in {\mathcal{C}(i)/j}}} {x_{c_p \to v_i}^{(l - 1)}}
\label{eq_v2c}
\end{equation}

while  the message from $c_j$ to $v_i$ is \ref{eq_c2v}
 
\begin{equation}
x_{c_j \to v_i}^{(l)} = 2{\tanh ^{ - 1}}\left( {\prod\limits_{\substack{\scriptstyle{v_q} \to c_j\\ \scriptstyle q \in {\mathcal{V}(j)/i}}} {\tanh \left( {\frac{{x_{v_q \to c_j}^{(l)}}}{2}} \right)} } \right) 
\label{eq_c2v}
\end{equation} 
where ${\mathcal{C}(i)/j}$ denotes all neighboring check nodes of $v_i$ excluding $c_j$, and ${\mathcal{V}(j)/i}$ denotes all neighboring variable nodes of $c_j$ excluding $v_i$,

\begin{equation}
x_{v_i}^{(l)} = {b_{v_i}} + \sum\limits_{\substack{c_p \to v_i\\p \in \mathcal{C}(i)}} {x_{{c_p} \to {v_i}}^{(l - 1)}}
\label{eq_bit_decision}
\end{equation} 

To circumvent the expensive computation of $\tanh$ function in \eqref{eq_c2v}, the MS  was  substituted for it via a simple approximation in \eqref{eq_ms}, at the cost of some performance loss.
\begin{equation}
x_{c_j \to v_i}^{(l)} =\left( \prod\limits_{\substack{\scriptstyle{v_q} \to c_j\\\scriptstyle {q \in v(j)/i}}} {sgn\left( x_{v_q \to c_j}^{(l)} \right)}\right) \mathop {\min }\limits_{\substack{\scriptstyle{v_q} \to {c_j}\\\scriptstyle {q} \in v(j)/i}} \left| {x_{v_q \to c_j}^{(l)}} \right|
\label{eq_ms}
\end{equation}

 What makes the MS more attractive is its characteristic of scale invariance, suitable for the case of channel noise with unknown $\sigma^2$. In comparison, the BP performance will suffer seriously with inaccurate estimation of $\sigma^2$ \cite{Lugosch2018b}.
 
To bridge the performance gap between MS and BP decoders, the NMS and OMS  were proposed to multiply or add  a constant correction term in the $min$ term of \eqref{eq_ms} respectively, the validity of which was verified in literature. 

\subsection{Related work of the NNMS}
Since performance of NOMS lags behind NNMS generally \cite{Lugosch2018b}, only the NNMS is discussed henceforth. 

A MS can be readily transformed into a trellis structure by unrolling each iteration of it. Then the NNMS as a special neural network, is born by adding  trainable parameters $\bm{\alpha},\bm{\beta},\bm{\gamma}$ as shown in  \ref{weighted_eq_v2c} and \ref{weighted_eq_c2v} on the edges of the trellis. These parameters, as a whole, will play a dual role of mitigating adverse effect due to the cycles in the Tanner graph and narrowing performance gap between the MS and BP. Till now one pending subject about the NNMS is the assignment of its parameters which dominates its performance.
\begin{equation}
	x_{v_i \to c_j}^{(l)}  = \alpha_{i}^{(l)}{b_{v_i}} + \sum\limits_{\substack{c_p \to v_i\\p \in {\mathcal{C}(i)/j}}} {\beta_{p,i}^{(l)}x_{c_p \to v_i}^{(l - 1)}}
	\label{weighted_eq_v2c}
\end{equation}
\begin{equation}
\resizebox{.9\hsize}{!}{$
	x_{c_j \to v_i}^{(l)} = \left(\prod\limits_{\substack{\scriptstyle{v_q} \to {c_j}\\\scriptstyle q \in v(j)/i}} {sgn\left( {x_{v_q \to c_j}^{(l)}} \right)}\right)\left(\gamma_{q,j}^{(l)} \mathop {\min }\limits_{\substack{\scriptstyle{v_q} \to c_j\\\scriptstyle q \in v(j)/i}} \left| {x_{v_q\to {c_j}}^{(l)}} \right|\right)
	$}
\label{weighted_eq_c2v}
\end{equation}
\renewcommand{\algorithmicrequire}{\textbf{Input:}}
\renewcommand{\algorithmicensure}{\textbf{Output:}}
\begin{algorithm}[h]
	\caption{NNMS decoding}
	\label{alg::message_flow_NNMS}
	\begin{algorithmic}[1]
			\Require
		channel signals $\bm{b}$, $\bm{H},T$, and well trained $\bm{\alpha},\bm{\beta},\bm{\gamma}$
		\Ensure
		estimated binary vector $\widehat{\mathbf{c}}$
		
		\State For any  $i\in\{1,2,...,N\}, j\in\{1,2,...,M\}$	
		\State  $x_{c_j \to v_i}^{(0)}=0,l=1$;		
	    \Repeat  
	\State calculate $v_i->c_j$ message with \ref{weighted_eq_v2c};  
	\State calculate $c_j->v_i$ message with \ref{weighted_eq_c2v};  
	\State $\widehat{\mathbf{c}}^{(l)}=(\hat{c}_1,\hat{c}_2,...,\hat{c}_N),\hat{c}_i=(1-sgn(x_{v_i}^{(l)}))/2$ by \ref{eq_bit_decision};

	\If {$\bm{H}\widehat{\mathbf{c}}^{(l)}=\mathbf{0}$ }
		\State return  $\widehat{\mathbf{c}}^{(l)}$;
	\Else
		\State $l= l+1$; 
	\EndIf 
	\Until{($l>T)$} 
	\State return  $\widehat{\mathbf{c}}^{(T)}$;
	\end{algorithmic} 
\end{algorithm}
 
 For the purpose of illustration, the  neural network architecture of NNMS is presented in Fig.~\ref{NNMS_structure}
\begin{figure}
\centering
        \def\pgfsysdriver{pgfsys -dvipdfmx.def}
        \usetikzlibrary{positioning, fit,calc,arrows.meta} 
        \tikzset{
        	block/.style={
        		draw, thick, text width=3cm, minimum height=1cm, align=center
        	}, 
        	line/.style={-latex},   
        	dash_box/.style ={
        		rectangle, 
        		rounded corners =5pt, 
        		minimum width =95pt, %
        		minimum height =218pt, %
        		outer sep=-5pt, 
        		draw=blue, 
        		densely dashed,
        		line width =1pt
        	},
        	omit_dash_box/.style ={
        	rectangle, %
        	minimum width =40pt, %
        	minimum height =200pt, %
        	outer sep=-5pt, %
        	},
        	horizontal_dash_box/.style ={
        	rectangle, %
        	rounded corners =5pt, %
        	minimum width =180pt, %
        	minimum height =20pt, %
        	outer sep=-1pt, %
        	draw=blue, %
        	densely dashed,
        	line width =1pt
        	},
        	arrow/.style = {
        		draw = purple, line width = 0.5pt, -{Latex[length = 2mm, width = 1.5mm]},
        	},
          rect1/.style = {
        	shape = rectangle,
        	draw = green,
        	text width = 3cm,
        	align = center,
        	minimum height = 1cm,
        	},
        	dline/.style ={color = blue, line width =2pt}
        }
        \pagestyle{empty}	
        \def\layersep{2.0cm}
\tikzset{global scale/.style={
    scale=#1,
    every node/.append style={scale=#1}
  }
}
    \begin{tikzpicture}[shorten >=1pt,->,draw=black!50,node distance=\layersep,global scale = 0.5]
		\tikzstyle{every pin edge}=[<-,shorten >=-1pt]
		\tikzstyle{neuron}=[circle,fill=black!25,minimum size=15pt,inner sep=0pt]
		\tikzstyle{input_neuron}=[circle,fill=black!100,minimum size=8pt,inner sep=0pt]
		\tikzstyle{variable_neuron}=[neuron, fill=green!50];		
		\tikzstyle{check_neuron}=[neuron, fill=blue!50];	
		\tikzstyle{output_neuron}=[neuron, fill=red!50];
			
		\tikzstyle{annot} = [text width=4em, text centered]
		\tikzstyle{annot1} = [text width=6em, text centered]		
		\foreach \name / \y in {1/1,2/2,3/3,4/4,5/5}
			\node[input_neuron, pin=below:\name] (Input-\y) at (\y+4.3,-9) {};	
		\node[input_neuron,pin=below:$N$] (Input-6) at (10.8,-9) {};	
		\node[annot]  at (10.1,-9) {$\cdots$};
		\node[annot]  at (10.1,-9.85) {$\cdots$};			
		\tikzstyle{every pin edge}=[<-,shorten >=2pt]
		\foreach \name / \y in {1/1,2/2,3/3,4/4,5/5}
			\node[variable_neuron,pin=below:] (I0v-\y) at (0,-\y) {};
		\node[annot] at (0,-5.75) {$\vdots$};
		\node[variable_neuron,pin=below:] (I0v-6) at (0,-6.5) {};	
		\foreach \name / \y in {1/1,2/2,3/3,4/4}
			\path[yshift=-0.5cm]
			node[check_neuron] (I1c-\y) at (\layersep,-\y) {};
		\node[annot] at (\layersep,-5.2) {$\vdots$};
		\node[check_neuron] (I1c-5) at (\layersep,-6) {};

		\foreach \name / \y in {1/1,2/2,3/3,4/4,5/5}
			\node[variable_neuron,right of=I1c,pin=below:] (I1v-\y) at (\layersep,-\y cm){};
		\node[annot]  at (2*\layersep,-5.75) {$\vdots$};
		\node[variable_neuron,pin=below:] (I1v-6) at (2*\layersep,-6.5) {};
		\foreach \name / \y in {1/1,2/2,3/3,4/4}
			\path[yshift=-0.5cm]
			node[check_neuron,right of=I1v] (I2c-\y) at (2*\layersep,-\y cm){};
		\node[annot] at (3*\layersep,-5.2) {$\vdots$};
		\node[check_neuron] (I2c-5) at (3*\layersep,-6) {};
		\foreach \name / \y in {1/1,2/2,3/3,4/4,5/5}
			\node[variable_neuron,right of=I2c,pin=below:] (I2v-\y) at (3*\layersep,-\y cm){};	
		\node[annot] at (4*\layersep,-5.75) {$\vdots$};
		\node[variable_neuron,pin=below:] (I2v-6) at (4*\layersep,-6.5) {};		
		\foreach \name / \y in {1/1,2/2,3/3,4/4}
		\path[yshift=-0.5cm]
		node[check_neuron,minimum size=5pt,fill=white!100,right of=I2v] (I3c-\y) at (3.75*\layersep,-\y cm){};
		\node[check_neuron,minimum size=5pt,fill=white!100] (I3c-5) at (4.75*\layersep,-6) {};
		\foreach \name / \y in {1/1,2/2,3/3,4/4,5/5}
		\node[variable_neuron,minimum size=5pt,fill=white!100,right of=I3c] (I3v-\y) at (4.25*\layersep,-\y cm){};	
		\node[variable_neuron,minimum size=5pt,fill=white!100] (I3v-6) at (5.25*\layersep,-6.5) {};			
		\foreach \name / \y in {1/1,2/2,3/3,4/4}
			\path[yshift=-0.5cm]
			node[check_neuron,right of=I2v] (Inc-\y) at (5*\layersep,-\y cm){};
		\node[annot] at (6*\layersep,-5.2) {$\vdots$};
		\node[check_neuron] (Inc-5) at (6*\layersep,-6) {};
		\foreach \name / \y in {1/1,2/2,3/3,4/4,5/5}
			\node[variable_neuron,right of=Inc,pin=below:] (Inv-\y) at (6*\layersep,-\y cm){};
		\node[annot] at (7*\layersep,-5.75) {$\vdots$};
		\node[variable_neuron,pin=below:] (Inv-6) at (7*\layersep,-6.5) {};							
				
		\foreach \name / \y in {1/1,2/2,3/3,4/4,5/5}
			\node[output_neuron,right of=I2v] (Output-\y) at (6.8*\layersep,-\y cm){};
		\node[annot,right of=I2v]  at (6.8*\layersep,-5.75){$\vdots$};	
		\node[output_neuron,right of=I2v] (Output-6) at (6.8*\layersep,-6.5 ){};		
		\foreach \y in {1,...,9}
			\node[annot](omit_\y) at (10.,-\y*0.65-0.5){$\cdots$};	
		\foreach \source in {1,3,5}
		\foreach \dest in {2,4}
			\path (I0v-\source) edge[-stealth] (I1c-\dest);
		\foreach \source in {2,4,6}
		\foreach \dest in {1,3,5}
			\path (I0v-\source) edge[-stealth] (I1c-\dest);	
		\foreach \source in {1,3,5}
		\foreach \dest in {2,4,6}
		 	\path (I1c-\source) edge[-stealth] (I1v-\dest);
		\foreach \source in {2,4}
		\foreach \dest in {1,3,5}
		  	\path (I1c-\source) edge[-stealth] (I1v-\dest);
		\foreach \source in {1,3,5}
		\foreach \dest in {2,4}
			\path (I1v-\source) edge[-stealth] (I2c-\dest);
		\foreach \source in {2,4,6}
		\foreach \dest in {1,3,5}
			\path (I1v-\source) edge[-stealth] (I2c-\dest);
		\foreach \source in {1,3,5}
		\foreach \dest in {2,4,6}
			\path (I2c-\source) edge[-stealth] (I2v-\dest);
		\foreach \source in {2,4}
		\foreach \dest in {1,3,5}
			\path (I2c-\source) edge[-stealth] (I2v-\dest);
		\foreach \source in {1,3,5}
		\foreach \dest in {2,4}
			\draw[-stealth] (I2v-\source) -- (I3c-\dest);
		\foreach \source in {2,4,6}
		\foreach \dest in {1,3,5}
			\draw[-stealth] (I2v-\source) -- (I3c-\dest);
		\foreach \source in {1,3,5}
		\foreach \dest in {2,4}
			\draw[-stealth] (I3v-\source) -- (Inc-\dest);
		\foreach \source in {2,4,6}
		\foreach \dest in {1,3,5}
			\draw[-stealth] (I3v-\source) -- (Inc-\dest);
		\foreach \source in {1,3,5}
		\foreach \dest in {2,4,6}
			\path (Inc-\source) edge[-stealth] (Inv-\dest);		
		\foreach \source in {2,4}
		\foreach \dest in {1,3,5}
			\path (Inc-\source) edge[-stealth] (Inv-\dest);	
		\foreach \source in {1,...,6}
		\path (Inv-\source) edge[-stealth] (Output-\source);
		
		\node[annot,above of = I0v-1, node distance=1cm] (h1) {\baselineskip=13pt  variable nodes\par};
		\node[annot,right of=h1] (h2) {\baselineskip=13pt check nodes\par };
		\node[annot,right of=h2] (h3){\baselineskip=13pt variable nodes \par};
		\node[annot,right of=h3] (h4){\baselineskip=13pt check nodes \par};
		\node[annot,right of=h4] (h5){\baselineskip=13pt variable nodes \par};
		\node[annot,above of = Inc-1, node distance=1.5cm] (h6){\baselineskip=13pt check nodes \par};
		\node[annot,right of=h6] (h7){\baselineskip=13pt variable nodes\par };
		\node[annot,above of = Output-1, node distance=1cm] {\baselineskip=13pt Soft output\par};
		\foreach \name / \y in {1/1,2/2,3/3,4/4,5/5}
			\node[annot,left of= I0v-\y,font=\footnotesize] (number\y) at (1.8,-\y-0.5){\y};
		\node[annot,left of= I0v-6,font=\footnotesize] (number6) at (1.8,-7){$N$};
		\node[annot,left of= I1v-6,font=\footnotesize] (number1_6) at (5.8,-7){$N$};	
		\node[annot,left of= I2v-6,font=\footnotesize] (number2_6) at (10.2,-7){$N$};	
		\node[annot,below = of I1c-5,font=\footnotesize,yshift=4cm] {$M$};	
		\node[annot,below = of I2c-5,font=\footnotesize,yshift=4cm] {$M$};
		\node[annot,below = of Inc-5,font=\footnotesize,yshift=4cm] {$M$};			
		\foreach \name / \y in {1/1,2/2,3/3,4/4,5/5}
		\node[annot,left of= Output-\y,font=\footnotesize] (soft_output\y) at (16.3,-\y-0.5){\y};
		\node[annot,left of= Output-6,font=\footnotesize] (soft_output6) at (16.3,-7){$N$};
		\node[dash_box] (first_iteration) at (3.05,-3.3) {};
		\node[annot1] (iteration_1) at (3.0,1.2) {1st iteration};	
		\node[dash_box] (second_iteration) at (7.1,-3.3) {};
		\node[annot1] (iteration_2)at (7.2,1.2){2nd iteration};	
		\node[omit_dash_box,align=center] (omit_iteration) at (10.15,-3.3) {};
		\node[annot1] (iteration_omit)at (10.15,1.2){$\cdots$};	

		\node[dash_box] (nth_iteration) at (13.1,-3.3) {};
		\node[annot1] (iteration_n)at (13.1,1.2){$T$th iteration};
		\node[horizontal_dash_box] (input_box) at(8.,-9) {};
		\node[annot,text width=8em,below of =input_box,align=center, node distance=40pt] (input_bits) {Soft input};	
		\draw[arrow](input_box) -- node[right,yshift = -5pt, xshift = 25pt]{$\bm{\alpha}^{(1)}$}(I0v-6);
		\draw[arrow](input_box) -- node[right,yshift = -5pt, xshift = 10pt]{$\bm{\alpha}^{(2)}$}(I1v-6);
		\draw[arrow](input_box) -- node[right,yshift = -5pt, xshift = -3pt]{$\bm{\alpha}^{(l)}$}(I2v-6);
		\draw[arrow,densely dotted](input_box) -- node[right,yshift = -5pt, xshift = -1pt]{$\cdots$}(9.6,-7);
		\draw[arrow](input_box) -- node[right,yshift = -5pt, xshift = -1pt]{$\bm{\alpha}^{(T)}$}(Inv-6);	
		\node[annot] at (1.,-1.){$\bm{\gamma}^{(1)}$};	
		\node[annot] at (3,-1.){$\bm{\beta}^{(1)}$};
		\node[annot] at (5.1,-1.){$\bm{\gamma}^{(2)}$};		
		\node[annot] at (7,-1.){$\bm{\beta}^{(2)}$};		
		\node[annot] at (9.1,-1.){$\bm{\gamma}^{(3)}$};		
		\node[annot] at (10.8,-.85){$\bm{\gamma}^{(T-1)}$};
		\node[annot] at (12.8,-1.){$\bm{\beta}^{(T)}$};				
	\end{tikzpicture}
\caption{A full-loaded NNMS framework in which the trainable parameters $\bm{\alpha},\bm{\beta},\bm{\gamma}$ assigned for each edge can be  shared each other or trimmed off to meet the need of applications, and the edge connections is in line with placement of non-zero elements of check matrix $\bm{H}$}
\label{NNMS_structure}
\end{figure}
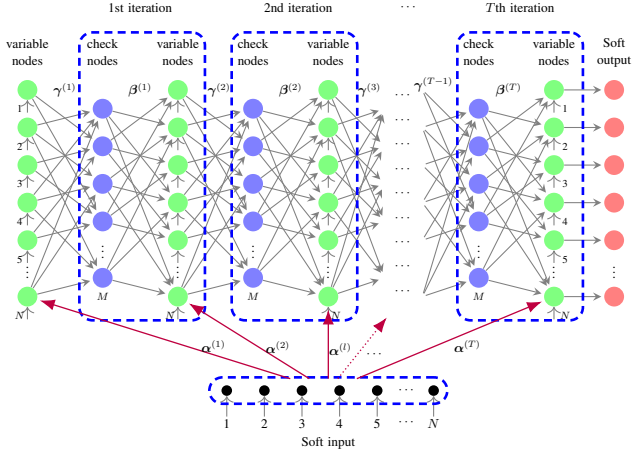
\begin{figure}[htbp]
	\centering
	\tikzstyle{startstop} = [rectangle, rounded corners, minimum width = 2cm, minimum height=1cm,text centered, draw = black, fill = red!40]
	\tikzstyle{io} = [trapezium, trapezium left angle=70, trapezium right angle=110, minimum width=2cm, minimum height=1cm, text centered, draw=black, fill = blue!40]
	\tikzstyle{process} = [rectangle, minimum width=3cm, minimum height=1cm, text centered, draw=black, fill = yellow!50]
	\tikzstyle{decision} = [diamond, aspect = 3, text centered, draw=black, fill = green!30]
	\tikzstyle{arrow} = [->,>=stealth]
	\tikzstyle{bag} = [align=center] 
	\begin{tikzpicture}[node distance=1.5cm,font=\fontsize{9}{9}\selectfont]
		\node (start) [startstop] {start};
		\node (pro1) [process, below of=start,bag] {Construct NNMS model,\\ with $\bm{\alpha},\bm{\beta},\bm{\gamma}$ initialized to be '1's};
		\node (pro2) [process, below of=pro1] {Define loss function for NNMS output};
		\node (pro3) [process, below of=pro2,bag] {Generate AWGN data batches,\\ Feeding NNMS model for training};
		\node (dec1) [decision, below of=pro3,yshift=-0.5cm,bag] {Loss value fixed\\
			or end of feeding};
		\node (pro4) [process, below of=dec1,yshift=-0.5cm] {Trained NNMS with optimized $\bm{\alpha},\bm{\beta},\bm{\gamma}$};
		\node (stop) [startstop, below of=pro4] {end};
		
		\draw [arrow](start) -- (pro1);
		\draw [arrow](pro1) -- (pro2);
		\draw [arrow](pro2) -- (pro3);
		\draw [arrow](pro3) -- (dec1);
		\draw [arrow](dec1) -- ($(dec1.east) + (0.5,0)$) node[anchor=north] {NO} |- (pro3);
		\draw [arrow](dec1) -- node[anchor=west] {YES} (pro4);
		\draw [arrow](pro4) -- (stop);	
	\end{tikzpicture}
\caption{Flow chart of training NNMS}
\label{fig: Flow chart of NNMS}
\end{figure}
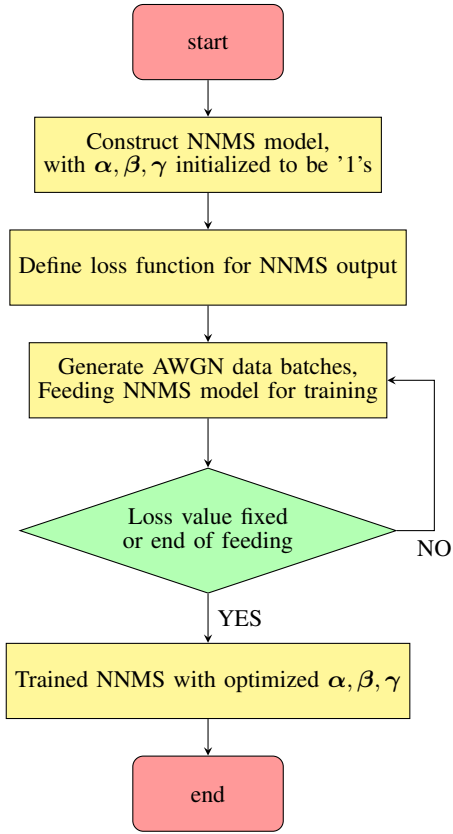

 In training phase, the neural decoders starts off with a data batch feeding, outputs an evaluation of the predefined loss function, then applies a gradient descent method to update all trainable parameters in a back propagation mode. More feedings, more updates. The training usually continues until the loss function halts  declining or all data epochs are iterated. A flow chart of training NNMS is shown in  Fig.~\ref{fig: Flow chart of NNMS}. 
 
 In testing phase, a well trained NNMS, as presented in  Alg.~\ref{alg::message_flow_NNMS}, is likely to improve decoding performance prominently.
\section{Motivation of our scheme}
\label{motivation}
Although we can generate a volume of AWGN random data of designated $\sigma^2$ without any effort, it is another story to determine a proper evaluation of  $\sigma^2$ for the training data of a specific code. As a result, when the input data with light $\sigma^2$ is less error-prone, the resulting NNMS can not cope with channel noisy signals even in medium SNR region. On the other, noisy data of aggravated $\sigma^2$ may prevent the NNMS from acquiring its decoding ability due to training divergence. 

Instead of the routine practice of blending the feeding data of some typical SNR points within a certain range \cite{gruber2017deep,Wang2020}, we propose to approximate it by generating training data according to a unique Gaussian density $\mathcal{N}(\mu_a,\sigma^2_a)$, deferred to be discussed in Section  \ref{simulations}.

In NNMS training phase, what intrigues us most is the relations between the loss function defined for decoder output and decoding metrics such as bit error rate (BER) or frame error rate (FER). However, it is too complex to analyze them in math expressions, we instead approach it by  tracing the training curves graphically in next section, hoping that will shed lights on what happened behind the scenes.

Naturally, given a well trained NNMS decoder, another question raised  is whether it is feasible to cut down the number of trainable parameters as many as possible so as to reach a better tradeoff between its decoding performance and computational complexity. After all, more trainable parameters imply more real multiplication operations involved. It is found that for some LDPC codes of rigorous algebraic structures or being randomly designed, at the cost of negligible performance loss, a full-loaded NNMS can be trimmed off even to retaining only one parameter which weighs the collection of incoming check messages and that of channel signals, suggesting that the importance of a proper placement of trainable parameters can not be overestimated.

\section{Experimental results}
\label{simulations}

Three LDPC codes of different designs will be discussed herein, consisting of  code A: a WiMAX (802.16) LDPC code (1056,880) \cite{Helmling2019}, code B: a finite geometry LDPC code (1023,781) \cite{Kou2001}, and code C: a Gallager  LDPC code (1008,504) \cite{Mackay2008}. For code A, it boasts of quasi-circular property, code B has high row redundancy due to its squared $\bm{H}$, while code C is randomly designed without any algebraic structure.

For the  ensemble of NNMS decoders, we can distinguish them from each other further by the number and locations of trainable parameters. The abbreviation 'SNNMS' refers to the neural decoder with a shared trainable parameter for each iteration at the check nodes side, that is, $\bm{\alpha}^{(l)}=1,\bm{\beta}^{(1)}=1,\bm{\gamma}^{(l)}=\gamma^{(l)}$. Likewise, 'UNNMS' is the one with a unique trainable parameter across all iterations, or $\bm{\alpha}^{(l)}=1,\bm{\beta}^{(1)}=1,\bm{\gamma}^{(l)}=\gamma$, while 'ANNMS' denotes a full-loaded NNMS decoder without any trimming as shown in Fig.~\ref{NNMS_structure}.

\subsection{Training phase}
We performed all the training and testing of the NNMS neural model of $T$ hidden layers on TensorFlow2.x of Colab or Kaggle cloud platforms.\footnote{Related source code will be open in github website after documented well}.

\subsubsection{Choice of loss function}
One prerequisite of training is to select a viable loss function among the ones off the shelf, on which we apply a stochastic gradient descent (SGD) method to optimize these trainable parameters. 

Given the authentic and estimated binary vectors $\bm{c}$ and $\bm{\widehat{c}}^{(j)},j=1,2,...T$, the hybrid loss function is defined  as follows, 
\begin{equation}
\label{def_loss}
\begin{aligned}
		\ell (\bm{c},\widehat {\bm{c}}) =  \rho\ell_{ce}(\bm{c},\widehat {\bm{c}})+ (1 - \rho )\kappa\ell_{mse}(\bm{c},\widehat {\bm{c}})
\end{aligned}
\end{equation}
where the weight factor $\rho=0.2$ and balance factor $\kappa=100$, and its cross entropy term is
\begin{equation}
\resizebox{.9\hsize}{!}{$
\label{ce_loss}
\begin{aligned}
	\ell_{ce} (\bm{c},\widehat {\bm{c}}) =  \frac{1}{{NT}}\sum\limits_{i = 1}^N \sum\limits_{j = 1}^T{\sum\limits_{z = 0}^1 {\left( {p({c_i} = z)\log \frac{1}{{p(\widehat c_i^{(j)} = z)}} } \right)} } 	
\end{aligned}
	$}
\nonumber 
\end{equation}
and the other mean squared error (MSE) term is
\begin{equation}
\resizebox{.9\hsize}{!}{$
	\label{mse_loss}
	\begin{aligned}
		\ell_{mse} (\bm{c},\widehat {\bm{c}}) =  \frac{1}{{N}}\sum\limits_{i = 1}^N {\sum\limits_{z = 0}^1}
		{p({c_i} = z){(p(\widehat c_i^{(T)} = z) - z)}^2}
	\end{aligned}
	$}
\nonumber 
\end{equation} 

The $\ell_{ce} (\bm{c},\widehat {\bm{c}})$, measures the density difference of the estimated codeword bits and the ground truth. Meanwhile, a 'multiloss' approach \cite{Nachmani2017}, via averaging the cross entropy of all $T$ iterations, ensures a stable outcome. For  $\ell_{mse} (\bm{c},\widehat {\bm{c}})$, it measures the average deviation of reliability of estimations from the original labels. This loss definition can effectively reduce the risk of training divergence caused by the oscillation of parameter updates.

\subsubsection{Generating of training data}
Notably, the assumption of all-zeros codeword transmitted in training, has no impact on the validness of the followed inferences, in the sense the trained NNMS  can deal with the cases of any codeword sending equally well,  This unique property, attributed to satisfying the message passing symmetry conditions \cite{Richardson2008}, greatly simplifies the training process. Correspondingly,  the loss definition \ref{def_loss} reduces to 
\begin{equation}
	\resizebox{.9\hsize}{!}{$
		\label{simplified_def_loss}		
		\ell (\bm{c},\widehat {\bm{c}}) =  \frac{\rho}{NT}\sum\limits_{i = 1}^N\sum\limits_{j = 1}^T 
		{ \log \frac{1}{p(\widehat c_i^{(j)}= 0)}} + \frac{(1 - \rho )\kappa}{{N}}\sum\limits_{i = 1}^N {p^2{(\widehat c_i^{(T)} = 0)}} 
		$}
\end{equation}

Although it is an easy task to simulate codeword sendings through communication channel disturbed by AWGN noise of designated $\sigma^2$,  we have to think twice about how to obtain volumes of 'qualified' data in feeding a NNMS, so that the training is well directed and the trained decoder has superior performance.  

In literature \cite{gruber2017deep,Wang2020}, each minibatch of training data comprises of samples drawn equally from a few evenly spaced SNR points. Notably, an inappropriate selection of these points may substantially affect training effectiveness. For one thing, the strong noise in harsh SNR region will ruin the manoeuvre of learning the structure of a neural decoder. Conversely, in extra high SNR region, the trained parameters may lack sufficient shifts, leading to the inability to decoding noisy signals. 

Assuming the blended minibatch consists of data from $I$ SNR points, then for each point, its component $Y_i\sim\mathcal{N}(1,\sigma_i^2)$, and the related LLR $Z_i=\frac{2}{{\sigma _i}^2}Y_i \sim \mathcal{N}(\frac{2}{{{\sigma _i}^2}},\frac{4}{{\sigma _i}^2})$, $i=1,2,...,I$. So a random variable $Z$ with the mixture density $f(Z)$, describing the blended data, is a weighted-sum of its component density  $f_{z_i}(Z)$, where $K$ is the number of information bits in a codeword.
\begin{equation}
\resizebox{.6\hsize}{!}{$
\begin{aligned}
	\quad \left(\text{SNR}\right)_i=\left( {\frac{{{E_b}}}{{{N_0}}}}\right)_i=10\log_{10}\frac{N}{{2K{\sigma_i ^2}}}
\end{aligned}
$}
\label{SNR_cal}
\end{equation}
\begin{equation}
\resizebox{.35\hsize}{!}{$
\begin{aligned}
	\quad f(Z) =\frac{1}{I}\sum\limits_{i = 1}^I {f_{Z_i}}({Z})
\end{aligned}
\label{mixture_density_cal}
$}
\end{equation}
\begin{equation}
\resizebox{.55\hsize}{!}{$
\begin{aligned}
	\mu_a&= E[Z] = \frac{1}{I}\sum\limits_{i = 1}^I E[{Z_i}] = \sum\limits_{i = 1}^I {\frac{2}{I\sigma_i ^2}}\\
	&=\frac{1}{\sigma_e-\sigma_s}\int_{\sigma_s}^{\sigma_e} {\frac{2}{x^2}}dx ,  {I\to+\infty}
\end{aligned}
\label{mean_cal}
$}
\end{equation}
\begin{equation}
\resizebox{.7\hsize}{!}{$
\begin{aligned}
{\sigma_a ^2}&=D[Z] = \frac{1}{I}\sum\limits_{i = 1}^I \left(({\frac{2}{\sigma_i ^2}})^2+{\frac{4}{\sigma_i ^2}}\right ) - {\mu_a ^2}\\
&=\frac{1}{\sigma_e-\sigma_s}\int_{\sigma_s}^{\sigma_e} {4(\frac{1}{x^4}+\frac{1}{x^2})}dx ,  {I\to+\infty}
\end{aligned}
\label{variance_cal}
$}
\end{equation}

\begin{figure}[htbp]
	\centering
	\centerline{\includegraphics[width=0.45\textwidth]{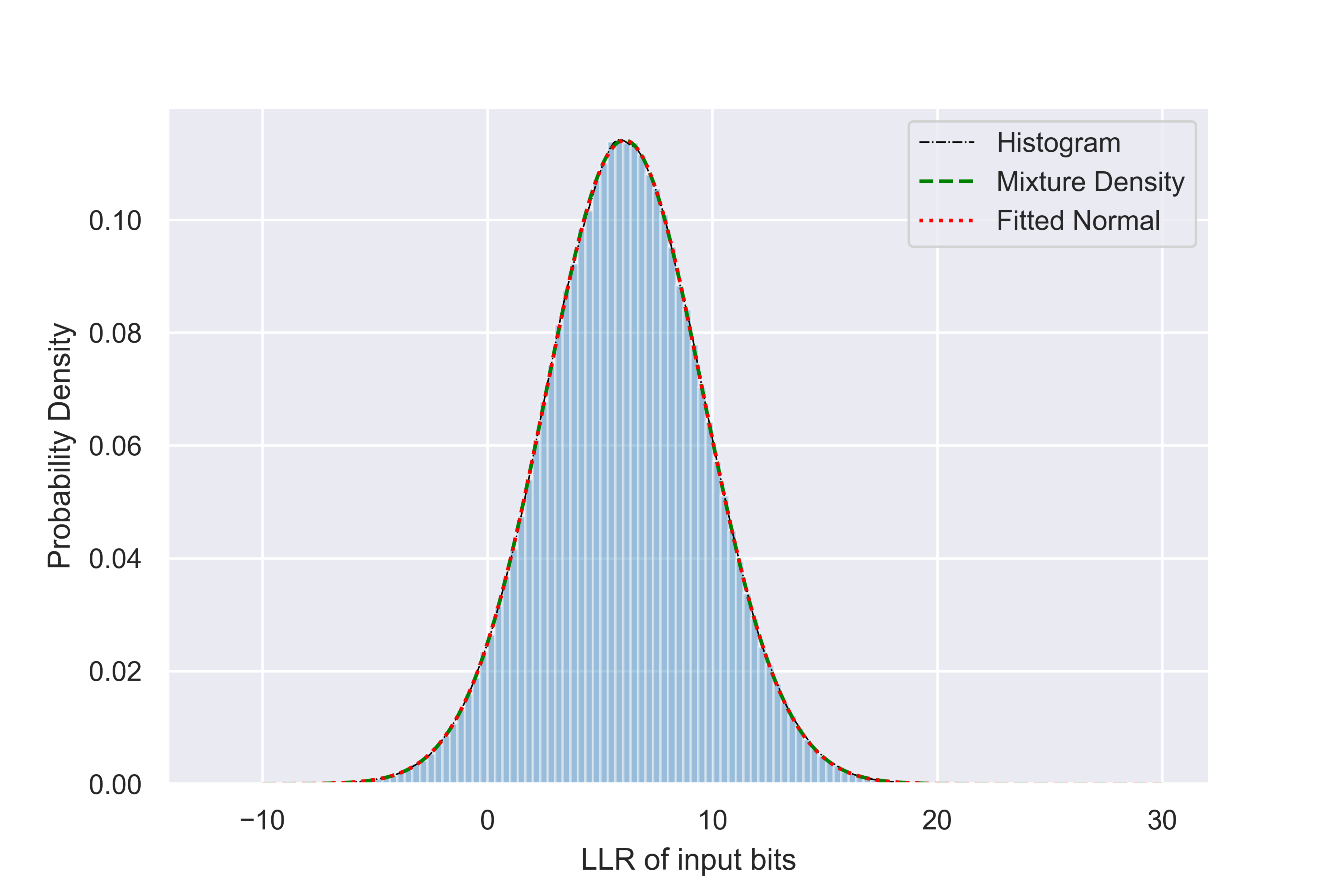}}
	\caption{Distribution of input' LLRs for (1023,781) code}
	\label{distribution_input}
\end{figure}

For code B, assuming SNR=$2.8\sim3.2$dB, $I=5$, with formulae  \ref{SNR_cal},\ref{mean_cal} and \ref{variance_cal}, $\mu_a=6.096$, $\sigma_a^2=12.232$. As shown in in Fig.~\ref{distribution_input},  the mixture density  perfectly matches a Monte-Carlo sampling  of the blended LLRs feeding a decoder. Furthermore, the fitting of a normal variable with the calculated  $(\mu_a,\sigma_a^2)$ almost overlays the underlying mixture density, which suggests that a minibatch of data may be approximately generated instead by directly sampling a normal variable of appropriate parameters. For the case of  ${I\to+\infty}$, we obtain the parameters $(\mu_a=6.093,\sigma_a^2=12.212)$, after identifying the corresponding standard deviation $\sigma_s$ and $\sigma_e$ of the interested SNR endpoints.

\subsubsection{Training settings}
For codes A,B,C, as listed in Table~\ref{tab_setting}, the training settings are similar up to minor shifts for some columns. 
\begin{table}[htbp]
	\setlength{\tabcolsep}{1.5mm}
	\caption{Training settings of Three codes}
	\begin{center}
	\resizebox{\linewidth}{!}{
	\begin{tabular}{|c|c|c|c|c|c|}
		\hline
		\textbf{Three}&{\textbf{SNR Range}}&\multicolumn{2}{|c|}{\textbf{Minibatch}}&\textbf{\# of}& \multirow{2}{*}{$T$}\\
			\cline{3-4} 
			\textbf{Codes} & (dB) &\textbf{\textit{number}} &\textbf{\textit{size}}&\textbf{epochs}&\\
			\hline
			code A&[3.2,3.8] &2000& 64 &6&20\\
			\hline
			code B&[2.8,3.2] &2000 &64 &6&10\\
			\hline
			code C&[1.8,2.4] &2000 & 32 &6&15/20\\
			\hline
		\end{tabular}
		}
		\label{tab_setting}
	\end{center}
\end{table}

All trainable parameters are initialized to be "1", which poses the starting point of a neural decoder to be a standard MS. Such a setting can expedite the training process to converge to a local optimum, compared with the routine normal distribution initialization. Meanwhile, we lean on the Adam optimizer  \cite{kingma2014adam} to update these parameters  with an exponential decaying learning rate wherein the initial rate is 0.002, decay rate 0.95 and decay steps 400.

\subsubsection{Training process analysis}
\begin{figure}[htbp]
	\centerline{\includegraphics[width=0.45\textwidth]{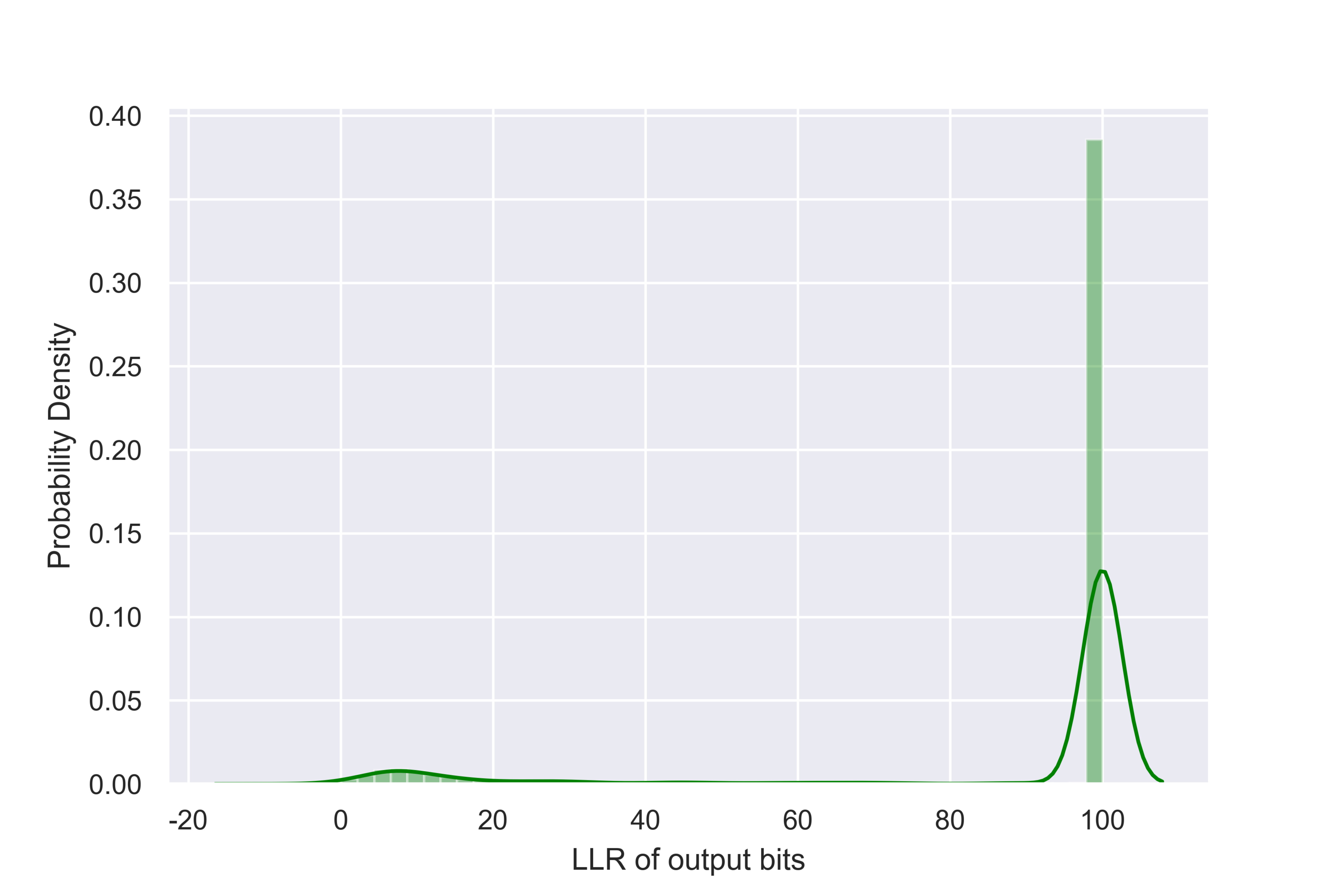}}
	\caption{Distribution of output' LLRs for (1023,781) code}
	\label{distribution_output}
\end{figure}

It is often difficult to get the exact soft decoder output distribution analytically. For instance of SNNMS decoding of code A, we resort to a Monte-Carlo sampling to obtain the fitted density curve in Fig.~\ref{distribution_output}. Apparently, the high density region concentrates on the positive clipped value 100, hinting  most bits made correct decision of being '0' with high confidence.

\begin{figure}[htbp]
\centering
	\centerline{\includegraphics[width=0.45\textwidth]{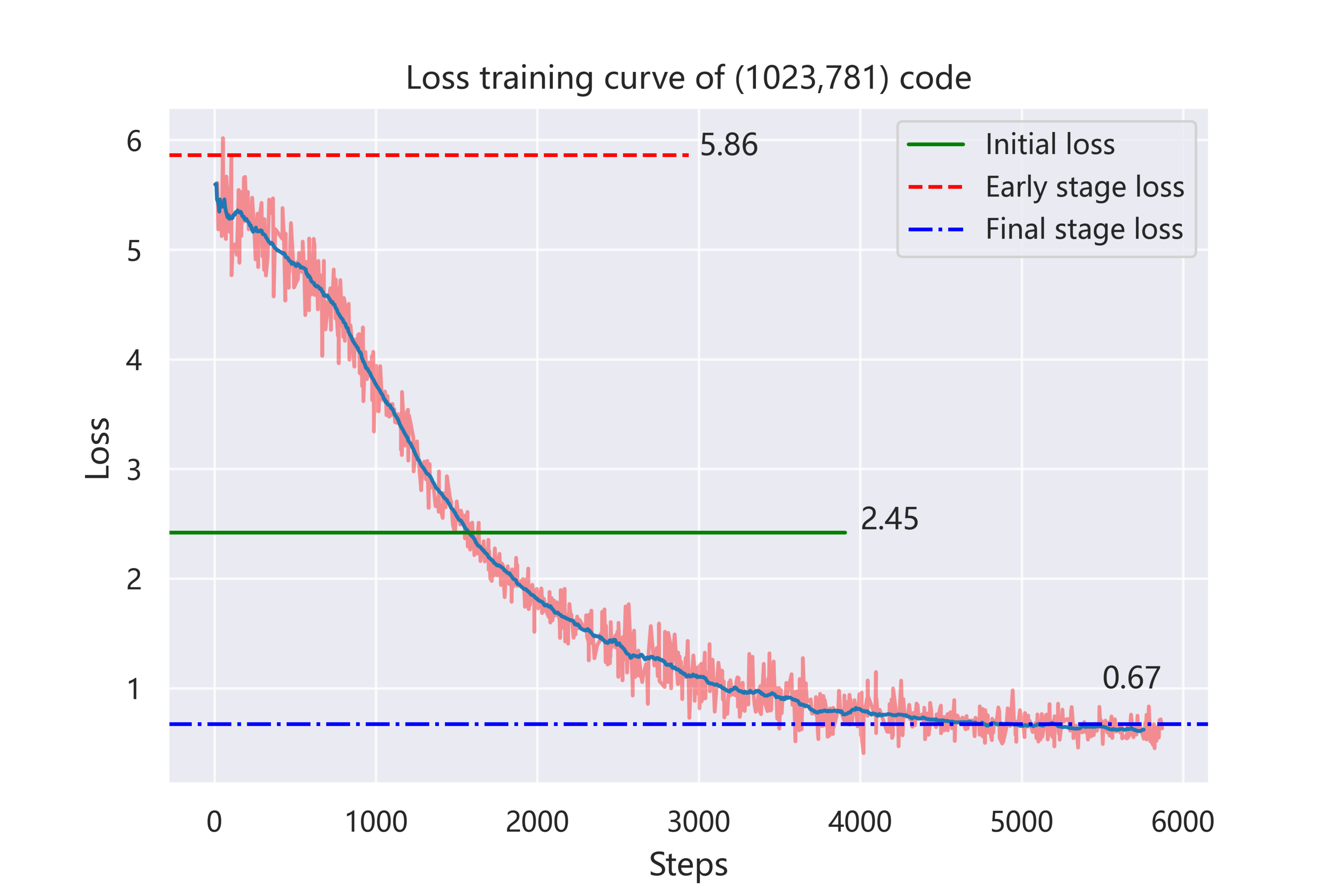}}
	\caption{Loss training curve of code B}
	\label{Loss_curve}
\end{figure}

\begin{figure}[htbp]
\centering
	\graphicspath{{./figs/}}
	\centerline{\includegraphics[width=0.45\textwidth]{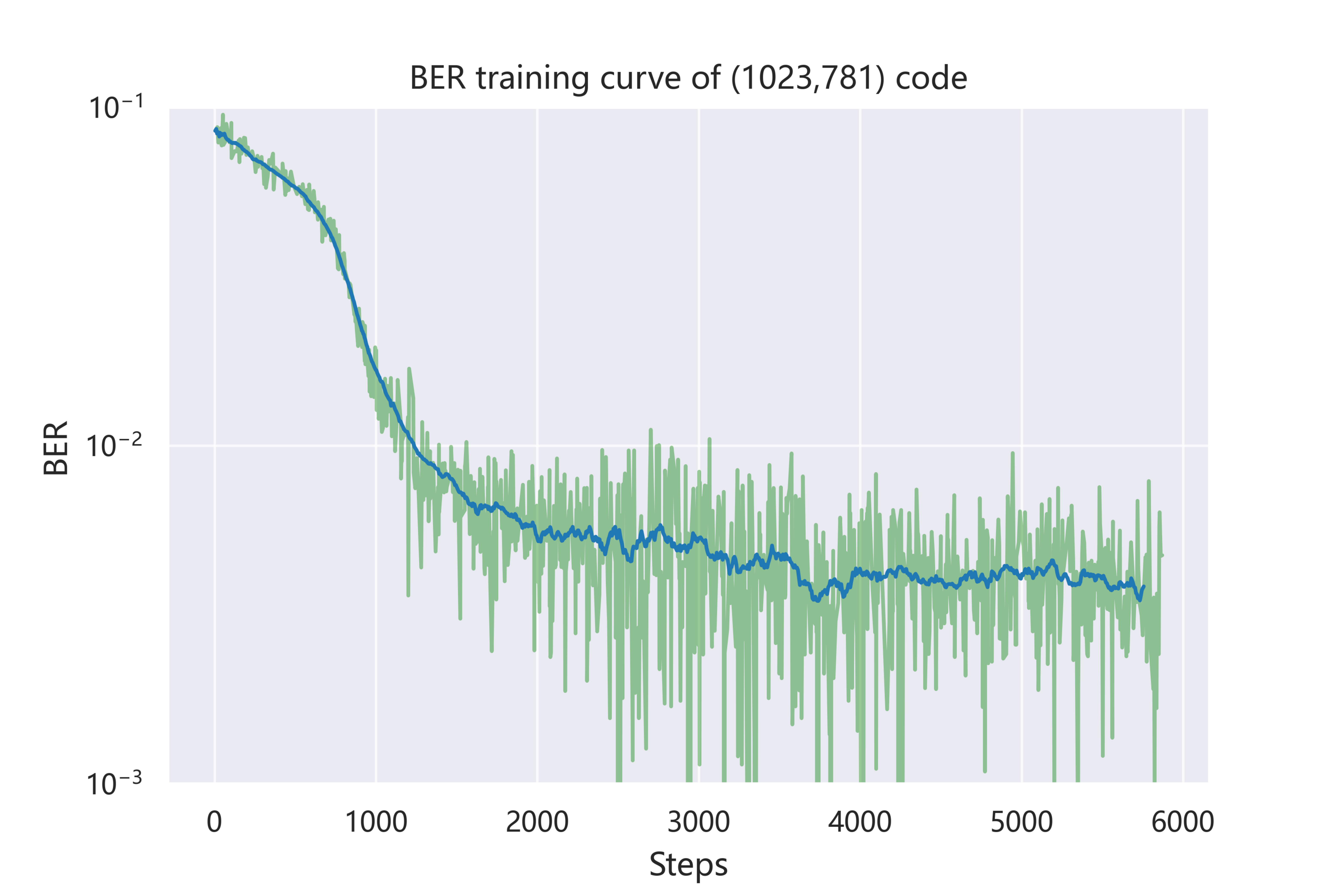}}
	\caption{BER training curve of code B}
	\label{BER_curve}
\end{figure}

\begin{figure}[htbp]
\centering
	\centerline{\includegraphics[width=0.45\textwidth]{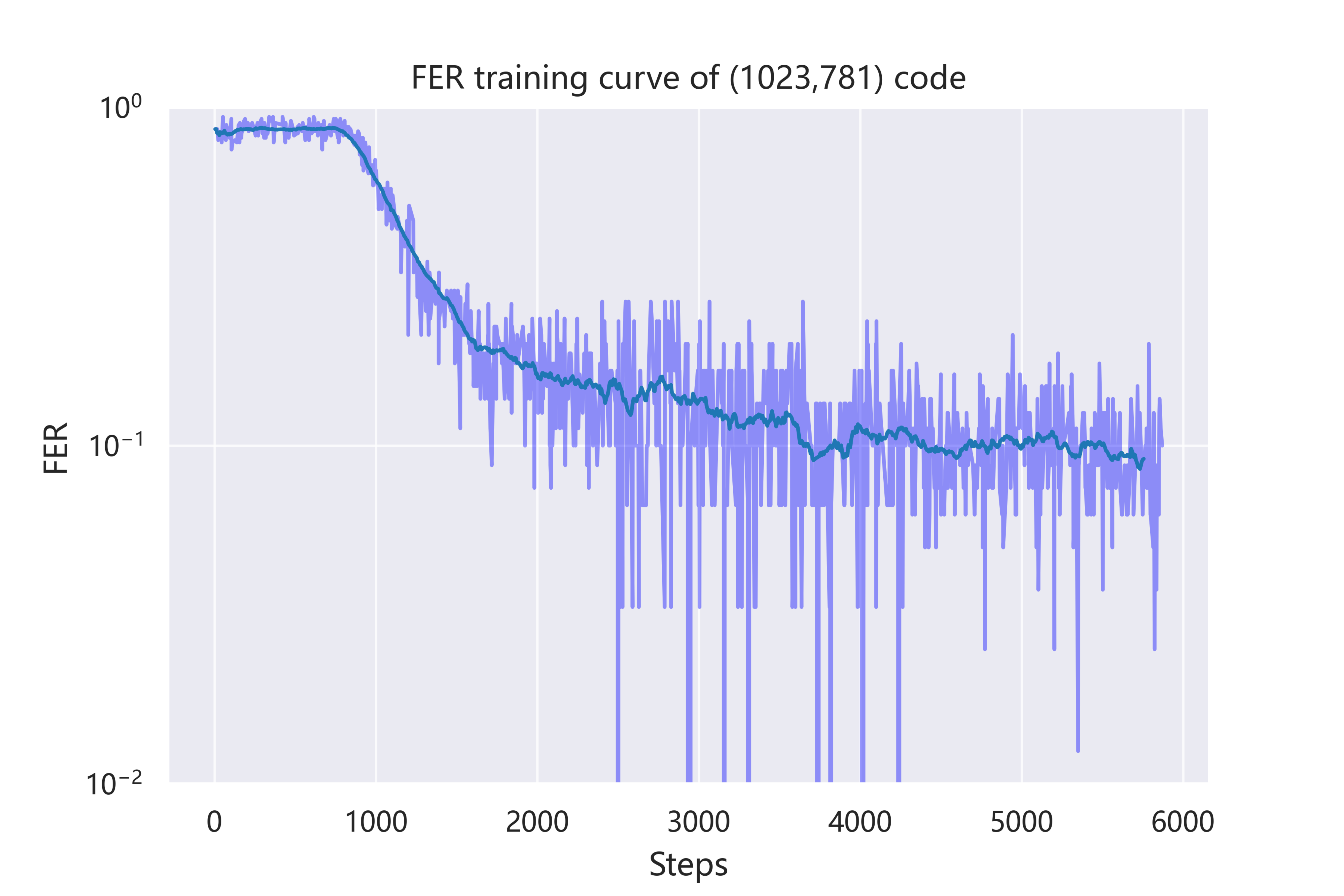}}
	\caption{FER training curve of code B}
	\label{FER_curve}
\end{figure}

Furthermore, to probe the connections between loss evaluation and  BER/FER, Figs.~\ref{Loss_curve}\ref{BER_curve}\ref{FER_curve} present three training evolution curves with/without smoothing.


 As shown in Fig.~\ref{Loss_curve}, Initially, the loss is evaluated with 2.45 for the input data, then it spikes to 5.86 after the feeding starts off, synchronized with a transitional BER/FER declining. Soon afterwards, with the proceeding of training process, the loss drops to 2.45 again, but with a substantially lowered BER/FER. 
 The training ends up with a loss of around 0.67. Meanwhile, the BER/FER improves steadily almost at the same pace with the loss curve, which reveals their strong positive correlation. 
 
 Specifically, calculated with 
\begin{equation}
\label{original_ber_cal}
\int_0^{ + \infty } {\frac{1}{{\sqrt {2\pi } \sigma_a }}} {e^{ - \frac{{{{(x - \mu_a)}^2}}}{{2{\sigma_a ^2}}}}}dx    
\end{equation}
where $\mu_a$ and $\sigma_a^2$ are acquired with \ref{mean_cal} and  \ref{variance_cal}, the initial BER of the input data is 0.04, at which moment the loss is 2.45 as marked in Fig.~\ref{Loss_curve}. When the loss curve revisits the horizontal line of 2.45, the accompanied BER is found  below 0.008 statistically. Consequently, an identical loss value is mapped to two different BERs, indicating the training process plays the role of polarization. That is, helped by message passing in the neural networking, most codeword bits at output can luckily make correct binary choice except for a few obstinate ones.  A similar observation holds for FER metric as well.

\subsubsection{Training result analysis}
For code B of SNNMS decoder, we traced the evaluation of its trainable parameters from beginning to end, and recorded some intermediate results in  Table~\ref{tab_weight_list}.

\begin{table}[htbp]
	\setlength{\tabcolsep}{1.5mm}
	\caption{Parameters evaluation and performance metrics for code B of SNNMS with $T=10$}
	\begin{center}
		\resizebox{\linewidth}{!}{
			\begin{tabular}{|c|c|c|c|c|}
				\hline
				\textbf{$i$-th step}&\textbf{weight list values}&\textbf{Loss}&\textbf{FER}&\textbf{BER}\\
				\hline
				0&\makecell[c]{0.54/  0.544/ 0.54/  0.544/ 0.544/ \\0.54/  0.54 / 0.54  0.54 / 0.54} &5.52&0.85&0.084\\
				\hline
				500&\makecell[c]{0.546  /0.562 / 0.548 / 0.509  /0.528/ \\ 0.51  / 0.289 -0.138/ -0.351/ -0.291}&5.04&0.91&0.052\\
				\hline
				1600&\makecell[c]{0.569 / 0.523/  0.418/ -0.074/ -1.021/\\ -1.435/ -1.13  -0.978/ -0.986/ -0.98} &2.79&0.23&0.0056\\
				\hline
				4400&\makecell[c]{0.307 /-0.051/ -0.974/ -1.748/ -1.177/\\ -1.06 / -1.034/ -1.028/ -1.065/ -1.126}&1.43&0.11&0.0042\\
				\hline
				11999&\makecell[c]{-1.059/ -1.153/ -1.114/ -1.085/ -1.094 /\\-1.072 -1.079 /-1.118 /-1.182/ -1.252}&0.49&0.075&0.0035\\
				\hline
			\end{tabular}
		}
		\label{tab_weight_list}
	\end{center}
\end{table}

Given an arbitrary input $x$, the softplus function $log(1+e^x)$ outputs a positive number, which fits well as a trainable parameter in our application. Hence, before training, all parameters are initialized to be '1'， implying  $x=0.542$, as observed in Table~\ref{tab_weight_list}, these parameters ends up training with a list whose elements are roughly '-1'. Considering the robustness of neural network, it naturally reminds us to attempt the traditional NMS using the softplus function evaluated at the point of '-1' as its correction factor, labelled as 'UNNMS' in the following tests. As illustrated in the next subsection,  such a simplification incurs no perceptible performance punishment. Furthermore, it is found weighting the LLR of each codeword bit with trainable  $\bm{\alpha}$ is invalid for the purpose of enhancing decoder performance, showing that the neural network treats every component of a codeword equally. 

\subsection{Testing phase}
A Monte-Carlo simulation is employed to accomplish the task of evaluating the performance of BER/FER for NNMS variants. For all codes A,B and C, at least 100 frame errors are required to be detected at each tested SNR point to reduce the estimation variance. And the maximum number of iteration $T$ is shown in the legends of each plot (cf. Table~\ref{tab_setting}). As benchmarks, the original plots of BP or other decodings are presented for comparison.

It has to be clarified there exists some performance discrepancy among different BP implementations\cite{RN2391}. It is roughly attributed to two variations. For one thing,  when the row weight $d_c$ of $\bm{H}$ is heavy, then the required successive multiplication of up to $d_c-1$ $tanh$ functions  in updating the outgoing message of each check node is a challenge in terms of  computation precision, and the situation is worsened in the scenario of all-zeros codeword sending in high SNR region where a clipping is  called. For another, in favor of parallel processing, our BP employs simple flooding messages scheduling instead of more advanced schedules \cite{zhang2002shuffled}. To avoid any ambiguity, our BP implementation is since referred as 'SBP'.

\subsubsection{Testing results analysis}
\label{test_result}
Assuming all neural decoders are available after training, given testing data feeding, SNNMS and UNNMS of code A with $T=20$ achieve almost identical performance with that of SBP with $T=40$, which is slightly behind the curve marked 'Msr-F(40)' originated in \cite{Helmling2019} for the reason of implementation details.
\begin{figure}[htbp]
	\centering
	\begin{tikzpicture}
		\begin{semilogyaxis}[
			scale = 0.75,
			xlabel={$E_b/N_0$(dB)},
			ylabel={BER/FER},
			xmin=2.0, xmax=4.5,
			ymin=1e-5, ymax=1,
			xtick={1.0,1.5,2,2.5,...,4.},
			legend pos = south west,
			ymajorgrids=true,
			xmajorgrids=true,
			grid style=dashed,
			legend style={font=\tiny},
			]
		\addplot[
		color=violet,
		mark=halfcircle*,
		]
		coordinates {
			(2.8,0.02229982112794613)
			(2.9,0.015013415404040407)
			(3.0,0.012101342667748916)
			(3.1,0.008381339436026937)
			(3.2,0.005912642045454545)
			(3.3,0.003989109848484849)
			(3.4,0.0022781356125914317)
			(3.5,0.00131227640993266)
			(3.6,0.0005800793328695114)
			(3.7,0.0002934823441021787)
			(3.8,0.00012546433243517654)
			(3.9,6.715446693171388e-05)
			(4.0,2.5327173764673786e-05)
		};	
		\addlegendentry{SNNMS-B(20)}
\addplot[
color=violet,
mark=halfcircle,
very thin
]
coordinates {
		(2.8,0.7395833333333334)
		(2.9,0.5338541666666666)
		(3.0,0.4575892857142857)
		(3.1,0.3506944444444444)
		(3.2,0.255)
		(3.3,0.18125)
		(3.4,0.10883620689655173)
		(3.5,0.07048611111111111)
		(3.6,0.03204719387755102)
		(3.7,0.017303719008264464)
		(3.8,0.008094394329896908)
		(3.9,0.004325929752066116)
		(4.0,0.0017683699324324324)
};	
\addlegendentry{SNNMS-F(20)}

\addplot[
color=purple,
mark=square*,
]
coordinates {
	(2.8,0.01978114478114478)
	(2.9,0.013622553661616162)
	(3.0,0.010644954004329006)
	(3.1,0.007606994423400672)
	(3.2,0.005271070075757576)
	(3.3,0.0035389510918003558)
	(3.4,0.0021179262996342736)
	(3.5,0.0011625299043062202)
	(3.6,0.000525272253787879)
	(3.7,0.0002663352272727272)
	(3.8,0.000115142163171773)
	(3.9,5.616710194260227e-05)
	(4.0,2.407611999625889e-05)
};
\addlegendentry{UNNMS-B(20)}
\addplot[
color=purple,
mark=square,
very thin
]
coordinates {
	(2.8,0.7395833333333334)
	(2.9,0.53125)
	(3.0,0.45535714285714285)
	(3.1,0.359375)
	(3.2,0.2575)
	(3.3,0.18566176470588236)
	(3.4,0.109375)
	(3.5,0.06611842105263158)
	(3.6,0.03204719387755102)
	(3.7,0.018153901734104045)
	(3.8,0.008675759668508287)
	(3.9,0.004055035506778566)
	(4.0,0.0019675925925925924)
};
\addlegendentry{UNNMS-F(20)}

		\addplot[
		color=blue,
		mark=diamond*,
		]
		coordinates {
		(2.8,0.01766098484848485)
		(2.9,0.010849461410984848)
		(3.0,0.009769324100378788)
		(3.1,0.006257739656177156)
		(3.2,0.003975880124777183)
		(3.3,0.002880366161616162)
		(3.4,0.0015305673009161377)
		(3.5,0.0009664163961038958)
		(3.6,0.0003862338098729229)
		(3.7,0.00017935032139577593)
		(3.8,9.832081006805091e-05)
		(3.9,3.847064393939394e-05)
		(4.0,2.0048643890225272e-05)
		};
	\addlegendentry{SBP-B(40)}
\addplot[
color=blue,
mark=diamond,
very thin
]
coordinates {

	(2.8,0.68125)
	(2.9,0.4296875)
	(3.0,0.421875)
	(3.1,0.2620192307692308)
	(3.2,0.18566176470588236)
	(3.3,0.13151041666666666)
	(3.4,0.07340116279069768)
	(3.5,0.045535714285714284)
	(3.6,0.02056451612903226)
	(3.7,0.010627104377104377)
	(3.8,0.006058061420345489)
	(3.9,0.00237312030075188)
	(4.0,0.0015757613579630554)
	
};
\addlegendentry{SBP-F(40)}

\addplot[
color=pink,
mark=triangle,
very thin
]
coordinates {
	(2.75,5e-1)
	(3.0,3e-1)	
	(3.25,9e-2)				
	(3.5,3e-2)	
	(3.75,4e-3)
	(4.0,1e-3)
	(4.25,1.5e-4)
};
\addlegendentry{Msr-F(40)}
		\end{semilogyaxis}
	\end{tikzpicture}
	\caption{BER/FER comparison for the decoding schemes of code A}
	\label{ber_fer_1056}
\end{figure}
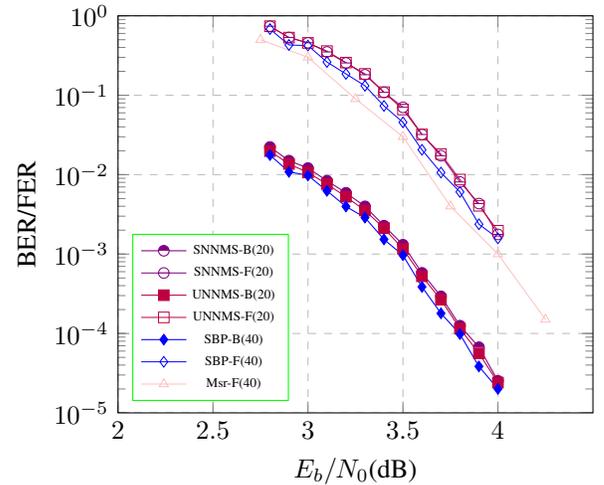

\begin{figure}[htbp]
	\centering
	\begin{tikzpicture}
		\begin{semilogyaxis}[
			scale = 0.75,
			xlabel={$E_b/N_0$(dB)},
			ylabel={BER/FER},
			xmin=1.5, xmax=4.,
			ymin=1e-6, ymax=1,
			xtick={1.0,1.5,2,2.5,...,4.},
			legend pos = south west,
			ymajorgrids=true,
			xmajorgrids=true,
			grid style=dashed,
			legend style={font=\tiny},
			]
			
			\addplot[
			color=purple,
			mark=square*,
			]
			coordinates {
				(2.5,0.02732160312805474)
				(2.6,0.022512218963831866)
				(2.7,0.014750733137829911)
				(2.8,0.008613554903877484)
				(2.9,0.005678152492668623)
				(3.0,0.002799975562072336)
				(3.1,0.0015988269794721408)
				(3.2,0.0005855938416422288)
				(3.3,0.00022500559422440508)
				(3.4,8.64431794223066e-05)
				(3.5,2.945824315503797e-05)
				(3.6,1.3440860215053762e-05)
			};
			\addlegendentry{UNNMS-B(10)}
			\addplot[
			color=purple,
			mark=square,
			very thin
			]
			coordinates {
				(2.5,0.54)
				(2.6,0.4566666666666666)
				(2.7,0.34)
				(2.8,0.19666666666666666)
				(2.9,0.13375)
				(3.0,0.06687500000000002)
				(3.1,0.04040000000000002)
				(3.2,0.015625000000000007)
				(3.3,0.0060240963855421725)
				(3.4,0.0024509803921568644)
				(3.5,0.000798722044728435)
				(3.6,0.0003500000000000002)
			};
			\addlegendentry{UNNMS-F(10)}
			
		\addplot[
		color=violet,
		mark=halfcircle*,
		]
		coordinates {
				(2.5,0.10807184750733138)
				(2.6,0.10310606060606062)
				(2.7,0.10371212121212121)
				(2.8,0.09930351906158356)
				(2.9,0.09536412512218963)
				(3.0,0.08796676441837732)
				(3.1,0.0857991202346041)
				(3.2,0.07287976539589444)
				(3.3,0.06457673509286413)
				(3.4,0.04977237815947494)
				(3.5,0.039705522971652)
				(3.5,0.039705522971652)
				(3.6,0.028025415444770285)
		};	
		\addlegendentry{MS-B(50)}
\addplot[
color=violet,
mark=halfcircle,
very thin
]
coordinates {
	(2.5,0.9975)
	(2.6,0.9875)
	(2.7,0.995)
	(2.8,0.9550000000000001)
	(2.9,0.925)
	(3.0,0.875)
	(3.1,0.8425)
	(3.2,0.736)
	(3.3,0.65)
	(3.4,0.5071428571428571)
	(3.5,0.4025000000000001)
	(3.5,0.4025000000000001)
	(3.6,0.2818181818181818)
};	
\addlegendentry{MS-F(50)}

		\addplot[
		color=blue,
		mark=diamond*,
		]
		coordinates {
				(2.5,0.01735483870967742)
				(2.6,0.011707163803937997)
				(2.7,0.006932551319648095)
				(2.8,0.004387202451718582)
				(2.9,0.00295227939216209)
				(3.0,0.0016543458781362005)
				(3.1,0.0011560018847092414)
				(3.2,0.0005866200973123773)
				(3.3,0.0003649615881700749)
				(3.4,0.00019799143228070408)
				(3.5,9.892473118279567e-05)
				(3.6,5.6129032258064523e-05)
		};
	\addlegendentry{SBP-B(50)}
\addplot[
color=blue,
mark=diamond,
very thin
]
coordinates {
	(2.5,0.32799999999999996)
	(2.6,0.22214285714285714)
	(2.7,0.135)
	(2.8,0.08216216216216217)
	(2.9,0.0549090909090909)
	(3.0,0.031770833333333325)
	(3.1,0.021726618705035946)
	(3.2,0.011235955056179782)
	(3.3,0.0067114093959731586)
	(3.4,0.003655006031363073)
	(3.5,0.0018733333333333243)
	(3.6,0.0010699999999999991)
	
};
\addlegendentry{SBP-F(50)}

		\addplot[
			color=pink,
			mark=triangle*,
			]
		coordinates {
				(2.0,5e-2)
				(2.2,3.5e-2)	
				(2.4,2.2e-2)				
				(2.6,1.2e-2)	
				(2.8,3.4e-3)
				(3.0,8e-4)
				(3.2,1.2e-4)
				(3.4,1.8e-5)
				(3.6,1.4e-6)
			};
		\addlegendentry{Msr-B(50)}

\addplot[
color=pink,
mark=triangle,
very thin
]
coordinates {
	(2.0,7e-1)
	(2.2,6e-1)	
	(2.4,4e-1)				
	(2.6,2.2e-1)	
	(2.8,6e-2)
	(3.0,0.018)
	(3.2,2.2e-3)
	(3.4,3.2e-4)
	(3.6,3e-5)
};
\addlegendentry{Msr-F(50)}
		\end{semilogyaxis}
	\end{tikzpicture}
	\caption{BER/FER comparison for the decoding schemes of code B}
	\label{ber_fer_1023}
\end{figure}
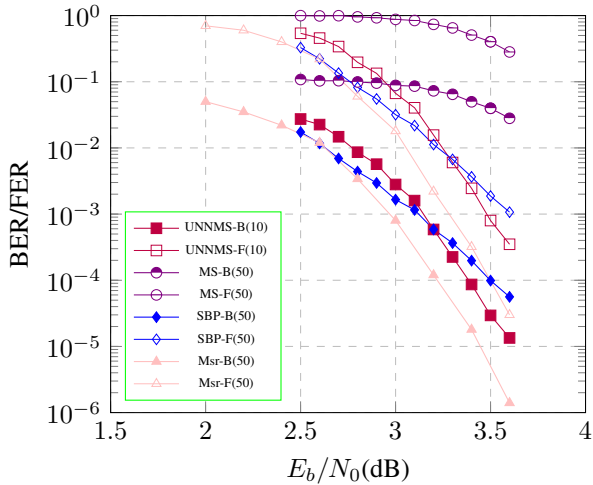

 For code B, Fig.~\ref{ber_fer_1023} shows BER/FER curves of various decoders, among which 'Msr-B/F (50)' denotes the original BER/FER plot  drawn in \cite{Kou2001} with $T=50$ for its BP. Since the performance of SNNMS or ANNMS leads UNNMS marginally, they are safely omitted in the plot. It is evident the MS is much less competitive compared with the others in all SNR region.  At the point BER = $10^{-4}$ of SNR waterfall region, UNNMS with $T=10$ surpasses SBP with  $T=50$, even though its performance lags behind the Msr-B(50) within $0.2$dB. The similar observation holds for the FER curves as well. 

\begin{figure}[htbp]
	\centering
	\begin{tikzpicture}
		\begin{semilogyaxis}[
			scale = 0.75,
			xlabel={$E_b/N_0$(dB)},
			ylabel={BER/FER},
			xmin=1, xmax=3.,
			ymin=1e-4, ymax=1,
			xtick={1.0,1.5,2,2.5,...,3.},
			legend pos = south west,
			ymajorgrids=true,
			xmajorgrids=true,
			grid style=dashed,
			legend style={font=\tiny},
			]

\addplot[
color=violet,
mark=halfcircle,
very thin
]
coordinates {
	(1.5,0.6221590909090909)
	(1.6,0.44166666666666665)
	(1.7,0.396484375)
	(1.8,0.24519230769230768)
	(1.9,0.18035714285714285)
	(2.0,0.11019736842105263)
	(2.1,0.06902472527472528)
	(2.2,0.03630780346820809)
	(2.3,0.02013221153846154)
	(2.4,0.010416666666666666)
	(2.5,0.004398634453781513)
};	
\addlegendentry{SNNMS-F(15)}

\addplot[
color=purple,
mark=square,
very thin
]
coordinates {
	(1.5,0.640625)
	(1.6,0.4479166666666667)
	(1.7,0.41796875)
	(1.8,0.25375)
	(1.9,0.19128787878787878)
	(2.0,0.11747685185185185)
	(2.1,0.07477678571428571)
	(2.2,0.03806818181818182)
	(2.3,0.02235320284697509)
	(2.4,0.011317567567567568)
	(2.5,0.0054009028374892515)
};
\addlegendentry{UNNMS-F(15)}

\addplot[
color=pink,
mark=triangle,
very thin
]
coordinates {
		(1.5,0.5795)
		(1.6,0.4777)
		(1.7,0.4121)
		(1.8,0.2853)
		(1.9,0.1932)
		(2.0,0.1181)
		(2.1,0.0761)
		(2.2,0.0408)
		(2.3,0.0230)
		(2.4,0.0131)
		(2.5,0.0052)
};
\addlegendentry{ANNMS-F(15)}
\addplot[
color=violet,
mark=halfcircle*,
]
coordinates {
	(1.5,0.4270833333333333)
	(1.6,0.3359375)
	(1.7,0.234375)
	(1.8,0.1609375)
	(1.9,0.10348360655737705)
	(2.0,0.06219059405940594)
	(2.1,0.034895833333333334)
	(2.2,0.01758356545961003)
	(2.3,0.006597951680672269)
	(2.4,0.0033333333333333335)
	(2.5,0.001234375)
};	
\addlegendentry{UNNMS-F(20)}
\addplot[
color=blue,
mark=diamond,
very thin
]
coordinates {
	(1.5,0.4354166666666667)
	(1.6,0.33223684210526316)
	(1.7,0.21458333333333332)
	(1.8,0.15320121951219512)
	(1.9,0.09421641791044776)
	(2.0,0.05844907407407408)
	(2.1,0.029628537735849055)
	(2.2,0.017351519337016574)
	(2.3,0.007329346557759626)
	(2.4,0.0035208333333333333)
	(2.5,0.001546875)
};
\addlegendentry{SNNMS-F(20)}

\addplot[
color=purple,
mark=square*,
]
coordinates {
	(2.0,0.02)
	(2.1,1e-2)
	(2.2,5e-3)
	(2.3,2.2e-3)
	(2.4,9e-4)
	(2.5,4e-04)
};
\addlegendentry{ADMM-F}

		\end{semilogyaxis}
	\end{tikzpicture}
	\caption{BER/FER comparison for the decoding schemes of code C}
	\label{fer_1008}
\end{figure}
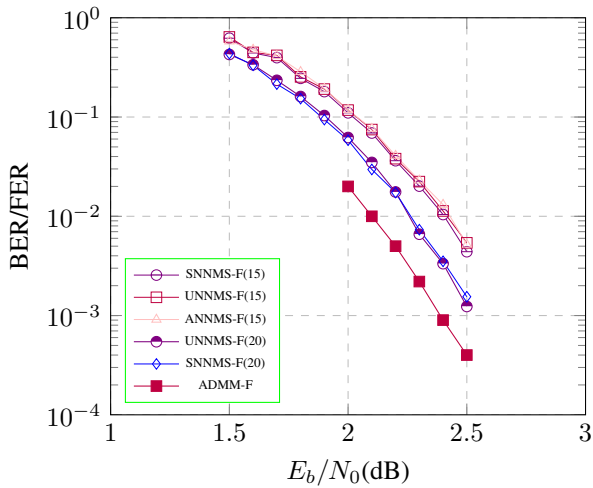

 For code C with $T=15,20$, Fig.~\ref{fer_1008} demonstrates that the performance of NNMS variants enhances with the increase of iterations $T$. Hopefully, with a further increase of $T$, it can come closer to the leading curve marked 'ADMM-F' (cf. \cite{RN2387}) which is enlightened by a differed decoding methodology.  

In sum, whether the codes are structured or randomly designed, the neural network manifests itself as a powerful alternative of designing low-complexity and high-performance decoders.

Last but not the least, we have to rethink why each member of the neural ensemble demonstrated a similar performance, regardless of a huge difference  in terms of number of trainable parameters. It is conjectured that for a code of algebraic circular or quasi-circular $\bm{H}$, its underlying encoding protects each bit or check of a codeword equally well, thus making the attempt of weighting per item in vain. On the other hand, for those codes without structure, the weighting functionality of a trainable parameter cancels off neutrally for all kinds of decoding cases.  With the fact that the weighting between collections, named 'sharing' in literature works well, as demonstrated for the case of UNNMS decoder, thus it reminds us to focus on its  location or impact scope when adding a trainable parameter, in case of a casual adding may lead to heavy training load and mean benefits.

\subsection{Complexity analysis}
For all three codes, compared with the other NNMS variant and BP, UNNMS is a competitive decoder, in the sense of achieving roughly equivalent performance with the minimum number of trainable parameters. Specifically, for code B,  UNNMS with the maximum number of iteration $T=10$ lags behind standard BP with $T=50$ within $0.2$dB at most. Besides that, the single multiplicative parameter of UNNMS can be approximated by bit shifting. Thus UNNMS actually requires only addition operations, while the BP is in need of volumes of multiplications besides the expensive $tanh$ functions. Likewise, for codes A and C, UNNMS  achieves equal performance compared with standard BP, at the cost of less or equal $T$ iterations. Therefore, simplified arithmetic operation, plus less iterations,  greatly promotes the throughput of UNNMS. 

Next we briefly compare the complexity of UNNMS with the popular alternating direction method of multipliers (ADMM) decoder \cite{RN2256}, which belongs to another class of LDPC decoding schemes rooted in mathematical programming. 

Given a code, assuming $d_c$ and $d_v$ are the average row weight and column weight of $\bm{H}$ respectively, the bit and check nodes update of UNNMS requires total $N(d_v+d_c)+2Md_c$ additions per iteration, according to \ref{eq_v2c} and \ref{eq_ms}. In comparison, $N(d_v+1)+3Md_c$ additions, $N$ multiplications and $N+M$ time-consuming projections are called per iteration  for the ADMM method. Considering the setting of $T$  of UNNMS is commonly fewer than that of ADMM method, we conclude that the former is much lighter in terms of complexity.

\section{Conclusions and future directions}
\label{conclusions}
In the framework of NNMS, we elaborated on how to generate high quality data as the feeding to the neural decoder in training. Then the roadmap of a training  process was investigated and two findings are worth mentioning. For one thing, there exists a strong positive correlation between loss function and BER/BER metrics after tracing the relevant evolution curves. For another, it is essential to investigate the structure of the LDPC codes to properly configure the placement of trainable parameters. The UNNMS, as a tailored NNMS, justified this assertion in extensive simulation by achieving a better tradeoff between performance and complexity.

For the loss definition of \ref{def_loss}, it is indicative of BER performance only, whereas a correct estimation of all codeword bits concurrently is ideal. To bypass this detour, it is helpful to update the loss definition by including some term containing FER information. It was reported in \cite{lugosch2018learning}  that the decoding performance was enhanced after exploiting a differentiable term pertained to decoding syndrome in the loss definition. For classical Reed–Muller codes, \cite{RN2359} designed a new loss definition for training  and proved its effectiveness in optimizing its survived parameters after reducing. It remains to be checked whether these definitions are applicable for more LDPC codes. 

It is likely the ensemble of neural decoders are robust to minor shifts of parameter evaluations. This feature can be exploited to approximate the arithmetic multiplication with a power of two, which is hardware friendly \cite{RN2311}.  Yet it has to be explored fully later. 

Besides that, it is always worth trying some novel networking structures for the neural decoders  to meet the demands of those challenging applications. 

\section*{Acknowledgement}
The authors would like to thank Google corporation for providing the excellent computing platforms of Colab and Kaggle online, which make it possible to train and test our models freely. Thanks also are given to the anonymous reviewers for providing their valuable feedback in improving the quality of this paper.

\end{document}